%% file: main.tex
\RequirePackage{amsmath}
\documentclass[runningheads,dvipdfmx]{llncs}

\usepackage[T1]{fontenc}
\usepackage{hyperref}
\usepackage{amsmath,amsfonts}
\usepackage[capitalize]{cleveref}
\usepackage{algorithm,algorithmic}
\usepackage{ulem}
\usepackage{multirow}
\usepackage {booktabs}
\usepackage{graphicx}
\usepackage{array}
\usepackage[dvipsnames]{xcolor}
\usepackage{tablefootnote}

\hypersetup{
    colorlinks=true,
    citecolor=Blue,
    linkcolor=black
}

\begin{document}

\title{Superpixel Attack}
\subtitle{Enhancing Black-box Adversarial Attack with Image-driven Division Areas}

\author{
    Issa Oe\inst{1}\orcidID{0009-0001-7216-2143} \and
    Keiichiro Yamamura\inst{1}\orcidID{0000-0003-4696-2881} \and \\
    Hiroki Ishikura\inst{1}\orcidID{0000-0002-4979-5276}\and
    Ryo Hamahira\inst{1} \orcidID{0009-0004-7634-451X} \and \\
    Katsuki Fujisawa\inst{2}\orcidID{0000-0001-8549-641X}
}
\authorrunning{I. Oe et al.}

\institute{
    Graduate School of Mathematics, Kyushu University \and
    Institute of Mathematics for Industry, Kyushu University \\
    \email{issa-oe@kyudai.jp}
}

\maketitle

\begin{abstract}
    Deep learning models are used in safety-critical tasks such as automated driving and face recognition. However, small perturbations in the model input can significantly change the predictions. Adversarial attacks are used to identify small perturbations that can lead to misclassifications. More powerful black-box adversarial attacks are required to develop more effective defenses. A promising approach to black-box adversarial attacks is to repeat the process of extracting a specific image area and changing the perturbations added to it. Existing attacks adopt simple rectangles as the areas where perturbations are changed in a single iteration. We propose applying superpixels instead, which achieve a good balance between color variance and compactness. We also propose a new search method, versatile search, and a novel attack method, Superpixel Attack, which applies superpixels and performs versatile search. Superpixel Attack improves attack success rates by an average of 2.10\% compared with existing attacks. Most models used in this study are robust against adversarial attacks, and this improvement is significant for black-box adversarial attacks. The code is avilable at \url{https://github.com/oe1307/SuperpixelAttack.git}.
\end{abstract}

\section{Introduction}

Deep learning models have recently found applications in automatic driving and face recognition tasks. These tasks are critical for safety, involving potential risks to life and information privacy. It has been observed that even small perturbations to the model input can significantly alter predictions \cite{adex_first}, leading to worst-case scenarios like accidents in automatic driving or information leakage in face recognition. Adversarial attacks are used to identify such perturbations that cause misclassifications. To counter these attacks, defense methods such as adversarial training \cite{training1,training2} and adversarial detection \cite{adv_detection1,adv_detection2} have been explored. However, more potent attacks are needed to develop more effective defenses.

This study targets black-box adversarial attacks, which operate under real-world constraints where only the model's predictions can be accessed. We focus on black-box adversarial attacks that aim to maximize attack success rates within allowed perturbations. A promising approach is to repeat the process of extracting a specific image area and changing the perturbations added to it.

Existing attacks use simple rectangles as the areas where perturbations are changed in a single iteration (\cref{update_area:previous_works}). However, it is natural to determine the areas based on the image's color information, as it directly influences the perturbation to be added. Therefore, we focus on the color variance of the area where perturbations are changed in a single iteration (\cref{update_area:color_variance}). Additionally, we focus on the compactness of the area, because existing attacks have adopted rectangles (\cref{update_area:compactness}). Through our analysis of the relationship among color variance, compactness, and attack success rates (\cref{update_area:superpixel}), we discovered that areas that are compact and have a low color variance result in higher attack success rates (\cref{update_area:experiment}). Consequently, we propose applying superpixels, which achieve a good balance between color variance and compactness.

Additionally, we introduce versatile search, a new search method that restricts the search to the boundary of perturbation and allows for searches using areas beyond rectangles. With these advancements, we propose Superpixel Attack, a novel attack method that applies superpixels and performs versatile search (\cref{superpixel_attack:update_area,superpixel_attack:versatile_search}). To evaluate the performance of Superpixel Attack, we conducted comparison experiments with existing attacks using 19 models trained on the ImageNet dataset \cite{imagenet} and available on RobustBench \cite{robustbench} (\cref{section:experiment}). Superpixel Attack significantly enhances attack success rates, resulting in an average improvement of 2.10\% compared to existing attacks. Considering that most models used in this study are robust against adversarial attacks, this improvement becomes especially noteworthy for black-box adversarial attacks. Our contributions can be summarized as follows:

\begin{enumerate}
    \item We analyze the relationship among the color variance, compactness, and attack success rates.
    \item We propose applying superpixels to black-box adversarial attacks and a new search method called versatile search.
    \item We conducted comparison experiments on Superpixel Attack, which applies superpixel and performs versatile search, and found improvement in attack success rates by an average of 2.10\% compared to existing attacks.
\end{enumerate}

\section{Preliminaries}

\subsection{Problem definition}

Let $H \in \mathbb{N}$ be the height, $W \in \mathbb{N}$ be the width, and $C \in \mathbb{N}$ be the number of color channels of the input image. Let $\mathcal{D} = [0, 1]^{H \times W \times C}$ denote the image space, $Y \in \mathbb{N}$ denote the number of classes of the model, and $f: \mathcal{D} \to [0, 1]^Y$ denote the classification model. The output of $f$ is the predicted probability of each class, and we denote $f_i(x) \in [0, 1]$ the predicted probability of class $i$ when image $x \in \mathcal{D}$ is the input. Adversarial attacks are to find an image $x_{adv} \in \mathcal{D}$ with the predicted label differs from the ground truth label $y \in \{1, \dots, Y\}$ of the original image $x_{org} \in \mathcal{D}$ by adding perturbations that are imperceptible to humans. The inputs generated by adversarial attacks are called adversarial examples. This study focuses on adversarial attacks that maximize attack success rates within the allowed perturbations. We set the allowed perturbation size $\epsilon \in \mathbb{R}^+$ and the loss function $L: [0, 1]^Y \times \{1, \dots, Y\} \to \mathbb{R}$, and solve the following constrained nonlinear optimization problem:
\begin{align}
    \label{formulation:adex}
    \begin{split}
        & \max_{x_{adv} \in \mathcal{D}} \quad L \left( f(x_{adv}), y \right)\\
        & \quad \textrm{s.t.} \qquad ||x_{adv} - x_{org}||_{\infty} \leq \epsilon
    \end{split}
\end{align}

\subsection{Related work}

Parsimonious attack \cite{parsimonious}, Square Attack \cite{square}, and SignHunter \cite{signhunter} have been proposed as black-box adversarial attacks defined by \cref{formulation:adex}. Parsimonious attack restricts the search space to the boundaries of allowed perturbations because attacks mostly succeed even on the boundaries. Square Attack achieves high success rates despite its reliance on random sampling. It is a part of AutoAttack \cite{AA}, a well-known white-box adversarial attack. SignHunter searches for adversarial examples by repeating image division and gradient direction estimation.

Black-box adversarial attacks that minimize perturbations under misclassification \cite{geoda,triangle} and those that reduce the number of perturbed pixels \cite{sparse_rs,saliency} have also been investigated. Attacks that generate adversarial examples from gradient information of surrogate models have also been proposed \cite{transferability1,transferability2}. These methods are based on transferability, that is, adversarial examples of one model often become those of others. However, training is required to make surrogate models resemble an attacking model and incurs high computational costs.

\section{Research on Update Areas}
\label{section:update_area}

\subsection{Update Areas of existing methods}
\label{update_area:previous_works}

The most promising approach for black-box adversarial attacks defined by \cref{formulation:adex} involves searching for adversarial examples by repeating the following steps: i. Extract a specific area from the image, ii. Collectively change the perturbation added to the extracted area, iii. Calculate the value of the loss function and update the perturbations when the loss increases. In this paper, we refer to the area where perturbations are changed in a single iteration as \textit{Update Area}. Existing black-box adversarial attacks have adopted simple rectangles as Update Areas. Parsimonious attack sets them using squares that divide the image equally. Square Attack sets them using randomly sampled squares from a uniform distribution. SignHunter sets them using rectangles that divide the image into equal horizontal sections.

\subsection{Color variance of Update Areas}
\label{update_area:color_variance}

As described in the previous section, Update Areas of the existing attacks are set using simple rectangles. However, it is natural to determine the area by considering the color information of the image because it determines the perturbation to be added. Therefore, we focus on the color variance of Update Areas. As a metric to express the color variance in divided areas of an image, Intra-Cluster Variation (ICV) \cite{icv} is proposed. ICV is calculated based on the following equation:
\begin{align}
    \textrm{ICV}
    = \frac{1}{\#\tilde{S}} \sum_{s \in \tilde{S}}{\frac{\sqrt{\sum_{p \in s}(I(p) - \mu(s))^2}}{|s|}},
\end{align}
where $\tilde{S}$ is the set of image segmentations. In this paper, it refers to the set of all Update Areas used in an attack. $s\in \tilde{S}$ denotes a single Update Area, and $p \in s$ denotes a pixel. $I(p)$ is the value of the pixel $p$ in the LAB color space\footnote{LAB color spaces in this paper refer to CIELAB (L, a*, b*) color space.} and $\mu(s)$ is the average value in the LAB color space within a single Update Area. $\#\tilde{S}$ is the number of Update Areas and $|s|$ is the number of pixels in a single Update Area. Smaller ICV indicates smaller color variance in each Update Area.

\subsection{Compactness of Update Areas}
\label{update_area:compactness}

Furthermore, considering that existing attacks use rectangles to set Update Areas, we focus on the compactness of Update Areas. The compactness (CO) \cite{compactness} is a metric calculated by dividing the size of the segments by that of a circle with the same perimeter length. The following equation defines this:
\begin{align}
    \textrm{CO} = \frac{\sum_{s \in \tilde{S}} Q(s) \cdot |s|}{\sum_{s \in \tilde{S}} |s|}, \qquad Q(s) = \frac{4 \pi |s|}{|R(s)|^2},
\end{align}
where $|R(s)|$ is the perimeter length of the Update Areas (number of pixels on the boundary). Higher CO indicates more centrally clustered Update Areas. We examined ICV and CO and attack success rates for various Update Areas construction in \cref{update_area:experiment}.

\subsection{Superpixel calculated by SLIC}
\label{update_area:superpixel}

\textit{Superpixel} is a set of pixels that are close in color and position. They have applications in object detection \cite{object_detection}, semantic segmentation \cite{semantic_segmentation}, and depth estimation \cite{depth_estimation}. Dong et al. proposed a white-box adversarial attack that adds the same perturbation to each superpixel to avoid disrupting the local smoothness of a natural image \cite{superpixel_white}. We use superpixels to improve the efficiency of black-box adversarial attacks. To the best of our knowledge, no black-box adversarial attacks that apply superpixels have been proposed. Various methods have been proposed for computing superpixels. We use one of the most popular methods: Simple Linear Iterative Clustering (SLIC) algorithm \cite{slic}. It places representative points at equal intervals according to the maximum number of segments and clusters pixels based on the k-means method. Let $(h_i, w_i)$ and $(h_j, w_j)$ be the positions in the image, and $(l_i, a_i, b_i)$ and $(l_j, a_j, b_j)$ be the values in the LAB color space. Clustering is performed based on similarity $k$.
\begin{align}
    \label{equation:slic}
    \begin{split}
        k_{color} = \sqrt{(l_i - l_j)^2 + (a_i - a_j)^2 + (b_i - b_j)^2} \\
        k_{space} = \sqrt{(h_i - h_j)^2 + (w_i - w_j)^2}                 \\
        k = \max(0,\ k_{color} + \alpha \cdot k_{space}),
    \end{split}
\end{align}
where $\alpha$ is a hyperparameter that weighs the positional distance relative to the color distance. $\alpha = 10$ is generally set to calculate superpixels. We examine the relationship between ICV, CO, and attack success rates for $\alpha = \pm 0.1$, $\pm 1$, $\pm 10$, $\pm 100$, $\pm 1000$ in \cref{update_area:experiment}. In addition, the SLIC implementation of scikit-image has the option to force each superpixel to be connected. The experiment in \cref{update_area:experiment} examine both cases. For $\alpha = 1000$, Update Areas are constructed as squares that divide the image equally, regardless of whether they are forced to be connected.

\subsection{Analysis of color variance and compactness}
\label{update_area:experiment}

\input{figure/co-icv}

The experiments use Salman et al. (ResNet-18) \cite{salman} trained on the ImageNet dataset and available on RobustBench. According to the RobustBench settings, we use 5,000 images randomly sampled from the ImageNet dataset, and the allowed perturbation size is set to $\epsilon = 4/255$. We adopted versatile search, a new search method proposed in \cref{superpixel_attack:versatile_search}. We examine attack success rates at the maximum iterations $T = 500$ for each Update Area construction. The attack success rate is calculated as follows: (number of misclassified images after the attack) / (total number of images), where the higher the attack success rate, the more powerful the attack. The seed value is fixed at 0. We used a CPU: Intel(R) Xeon(R) Gold 5220R CPU@2.20GHz$\times$2, GPU: Nvidia RTX A6000, RAM:768GB. The results are shown in \cref{figure:co-icv}.

Each point in \cref{figure:co-icv} represents the values of CO and ICV for different Update Area construction. The numerical values represent the attack success rate at the point. The horizontal axis represents the value of CO, and the right side indicates that more centrally clustered Update Areas are constructed. The vertical axis represents the value of ICV, where the upper side indicates that Update Areas with lower color variance are constructed. Note that the same Update Areas are constructed for some parameters of $\alpha$, and the points with equal ICV, CO, and attack success rates coincided with each other. For some representative points, the Update Areas generated by the SLIC algorithm are shown in different colors. This result indicates that it is effective to set Update Areas that are compact and have a low color variance.

\section{Superpixel Attack}
\label{section:superpixel_attack}

Based on the analysis in \cref{section:update_area}, we consider applying superpixels, which achieve a good balance between color variance and compactness, to black-box adversarial attacks. In this section, we describe the construction of Update Areas using superpixels (\cref{superpixel_attack:update_area}) and a new search method called versatile search (\cref{superpixel_attack:versatile_search}). We propose a novel attack method called \textit{Superpixel Attack} that sets Update Areas using superpixels and performs versatile search. An overview of Superpixel Attack is shown in \cref{figure:superpixel_attack}, and the pseudo-code is shown in \cref{algorithm:superpixel_attack}.

\input{figure/superpixel_attack}

\subsection{Update Areas using superpixels}
\label{superpixel_attack:update_area}

Below, we describe the construction of Update Areas using superpixels. Inspired by existing attacks, Update Areas are set using a few segments of superpixels at an early stage and many segments of superpixels as the attack progresses. Specifically, the segment ratio $r$ is given and superpixels $\mathcal{S}$ are computed following the maximum number of segmentations $n = r^j$ $(j = 1, 2, \dots)$. Let $S$ be the set of Update Areas constructed for each maximum number of segments $n$. The original image $x_{org}$ is divided into superpixels $\mathcal{S}$ for each RGB color channel $\{1, \dots, C\}$, which are set as Update Areas $S = \mathcal{S} \times \{1, \dots, C\}$. Note that the maximum number of superpixel segments $n$ is not always equal to the number of superpixels computed $\#\mathcal{S}$ in the SLIC algorithm employed in this study. The segment ratio is set to $r=4$ based on pre-examination. We set $\alpha = 10$ and force the areas to be connected.

\subsection{Procedure of versatile search}
\label{superpixel_attack:versatile_search}

Below, we describe a new search method called \textit{versatile search}. It searches only the boundaries of the allowed perturbations $\{-\epsilon, \epsilon\}^{H \times W \times C}$ according to the analysis by Moon et al. \cite{parsimonious}. At the beginning of the search, the perturbations are initialized with $\mathcal{E}_{best} = \{\epsilon\}^{H \times W \times C}$. Let $\mathcal{A}$ be the entire area of the image and initialize the set of Update Areas with $S=\{\mathcal{A}\}$. The best loss is initialized as $\mathcal{L}_{best}=-\infty$. The following steps are repeated until the number of iterations $t$ reaches the maximum iterations $T$.

First, the next area where the perturbations are changed is randomly extracted $s\in S$. In the first iteration, Update Area is set to the entire image ($s=\mathcal{A}$). Only the perturbations in the extracted Update Area $\mathcal{E}_{best}[s]$ is flipped to generate new perturbations $\mathcal{E}$. These perturbations $\mathcal{E}$ are added to the original image $x_{org}$, and the loss $\mathcal{L}$ is calculated. When the calculated loss $\mathcal{L}$ is higher than the best loss $\mathcal{L}_{best}$, the best loss $\mathcal{L}_{best}$ and the perturbation $\mathcal{E}_{best}$ are updated. Superpixels are computed when all Update Areas are searched ($S = \emptyset$), and new Update Areas are set using them.

When the attack is completed, the image with the best loss $x_{best}$ is returned. Superpixel Attack employs CW loss \cite{cw_loss} ($L_{cw}$) as the loss function based on pre-examination. CW loss is calculated as follows:
\begin{align}
    L_{cw}(f(x), y) = \max_{{i \neq y}}f_i(x) - f_y(x)
\end{align}

\section{Experiments}
\label{section:experiment}

In this section, we describe the comparison experiments conducted to confirm the performance of Superpixel Attack. We compare it to Parsimonious attack (Parsimon) \cite{parsimonious}, Square Attack (Square) \cite{square}, SignHunter (SignH) \cite{signhunter}, and Accelerated SignHunter (AccSignH) \cite{acc_signhunter} as a baseline. All of these are black-box adversarial attacks with the same problem settings. The experiments use 19 models trained on the ImageNet dataset and available on RobustBench. According to the RobustBench settings, we use 5,000 images randomly sampled from the ImageNet dataset, and the allowed perturbation size is set to $\epsilon = 4/255$. We examine the attack success rates at the maximum iterations $T = 100$ and $1000$. The baseline hyperparameters are the same as those in the original paper. The seed value is fixed at 0. We use the same computational environment as in \cref{update_area:experiment}. \cref{table:result} presents the results. The highest attack success rate for each iteration is bolded, and the difference between the best baseline method and Superpixel Attack is noted on the right side.

\input{table/result}

The results in \cref{table:result} show that Superpixel Attack improves the attack success rates by an average of 1.65\% for 100 iterations and 2.10\% for 1000 iterations compared to existing attacks. Most models used in this study are robust against adversarial attacks, and this improvement is significant for black-box adversarial attacks. In fact, the difference between the second-best and next-best existing attacks averaged 0.67\% for 100 iterations and 0.71\% for 1000 iterations. For Wong (ResNet-50), PyTorch (ResNet-50), and Singh (ViT-S+ConvStem), we plot the trends of attack success rates per iteration for each attack method in \cref{figure:plot_acc}.

\input{figure/plot_acc}

\cref{figure:plot_acc} indicates that Superpixel Attack achieves high success rates in all iterations, including the PyTorch (ResNet-50) model in contrast to SignHunter, which only has high success rates in short iterations. For the other models, each attack method exhibited trends similar to those of Wong (ResNet-50) and Singh (ViT-S+ConvStem). Furthermore, \cref{figure:computational_time} shows the computational time for superpixels and forward propagation in Superpixel Attack. Although it depends on the computational environment, the computation time of superpixels is less than that of the forward propagation. This indicates that applying superpixels to adversarial attacks is practical in terms of the computation time. For 1000 iterations, the superpixel computation accounts for a very small percentage of the attacks, as indicated by the orange bars.

\input{figure/computational_time}

\section{Conclusion}

This study demonstrated that the attack success rates are related to the color variance and compactness of the Update Area. The experimental results suggest that Update Areas with low color variance and high compactness is desirable. Therefore, we propose the Superpixel Attack, which employs superpixels as Update Areas to achieve a good balance between color variance and compactness. The comparison experiments show that the Superpixel Attack improves the attack success rates by an average of 2.10\% compared with existing methods for 1000 iterations, which is significant for black-box adversarial attacks. This study indicates that adjusting the Update Areas according to the image can enhance the attack success rates.

\textbf{Acknowledgements}\ 
This research project was supported by the Japan Science and Technology Agency (JST), the Core Research of Evolutionary Science and Technology (CREST), the Center of Innovation Science and Technology based Radical Innovation and Entrepreneurship Program (COI Program), JSPS KAKENHI Grant Number JP16H01707 and JP21H04599, Japan.

\bibliographystyle{splncs04}
\bibliography{reference}

\end{document}

%% file: figure/co-icv.tex
\begin{figure}[htbp]
    \centering
    \includegraphics[width=\linewidth]{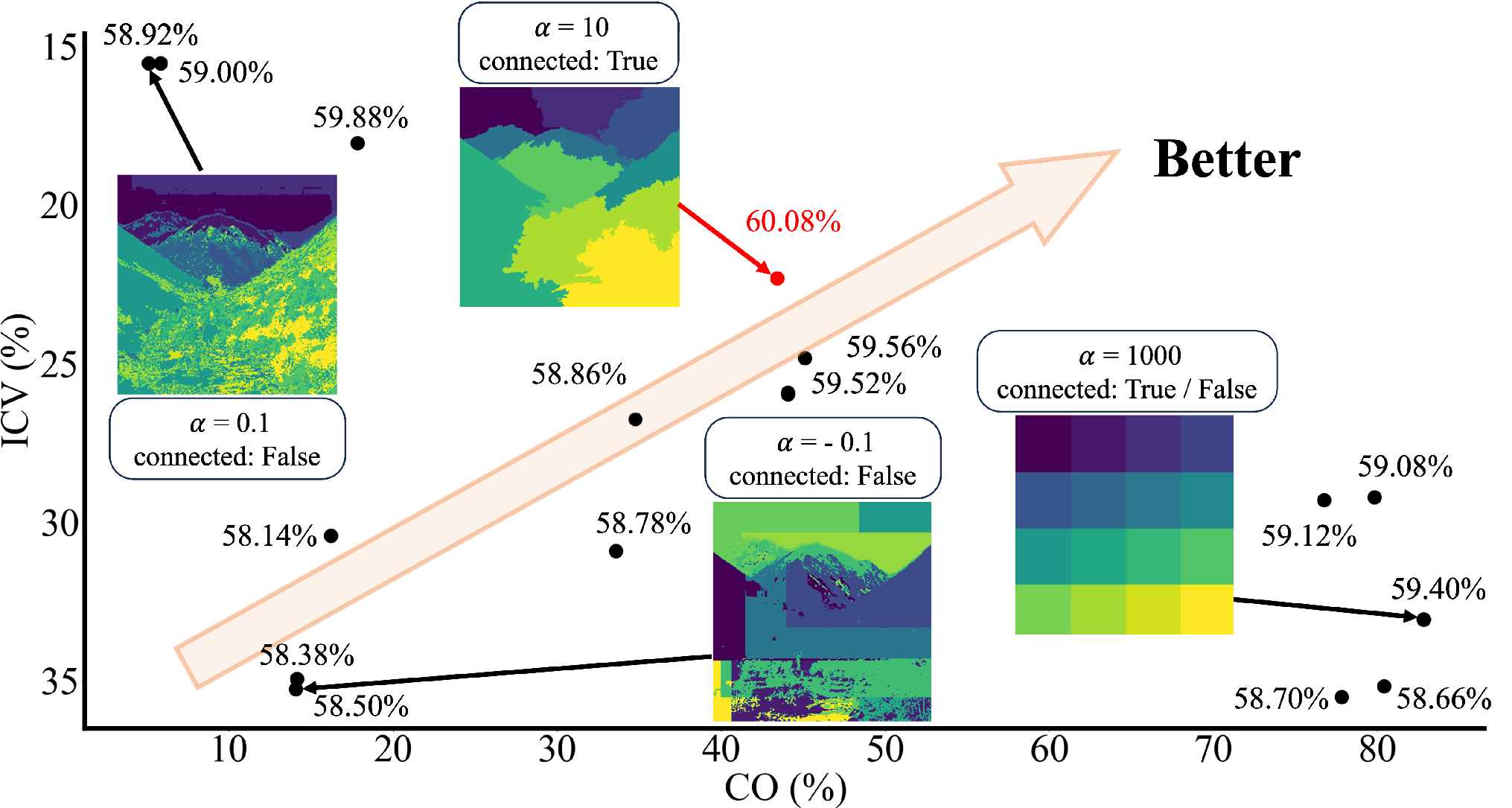}
    \caption{Relationship between ICV, CO and attack success rates}
    \label{figure:co-icv}
\end{figure}

%% file: figure/superpixel_attack.tex
\begin{figure}[htbp]

    \begin{minipage}{\linewidth}
        \centering
        \includegraphics[width=\linewidth]{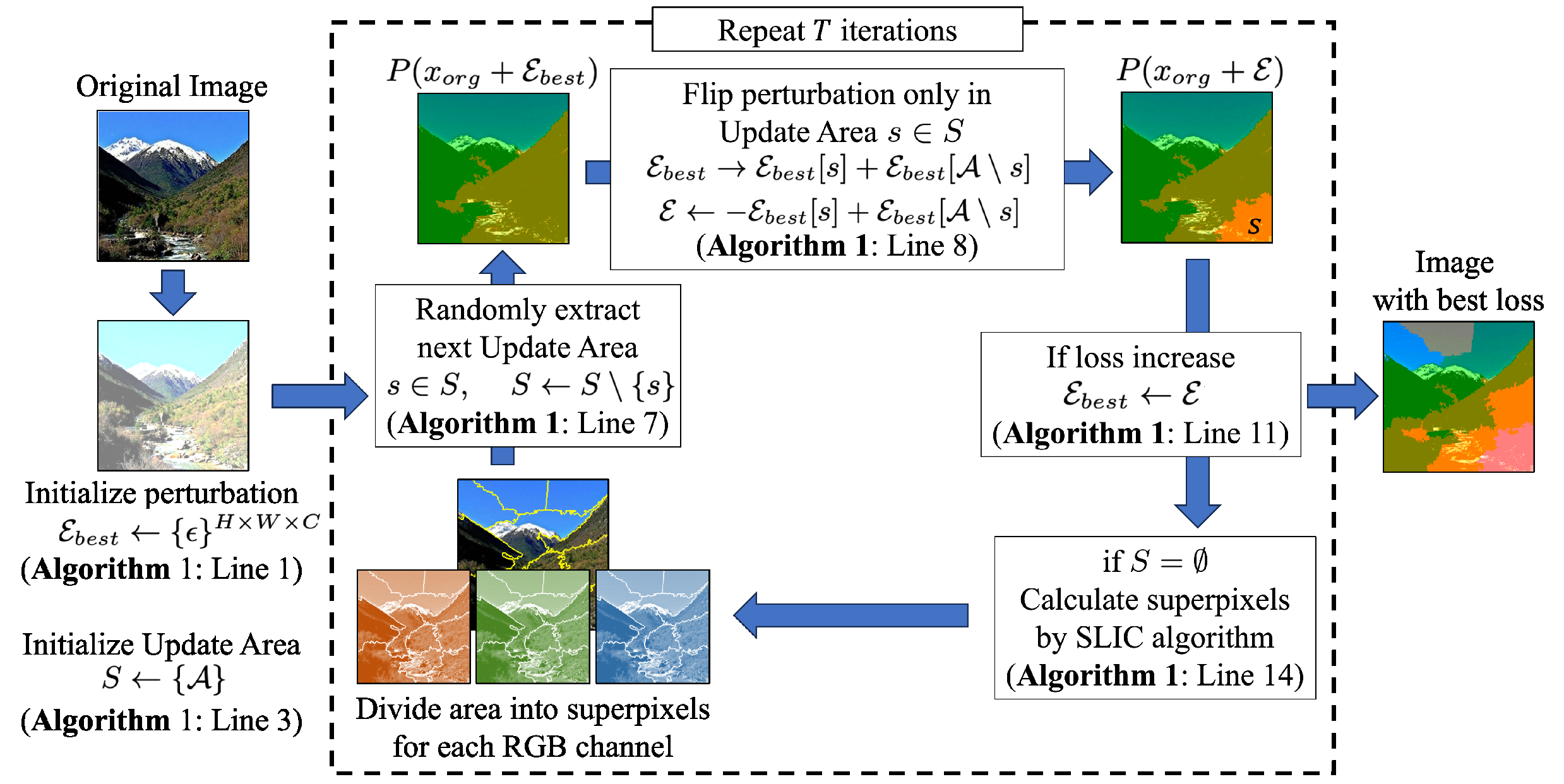}
        \caption{Flow of proposed method: Superpixel Attack}
        \label{figure:superpixel_attack}
    \end{minipage}

    \begin{minipage}{\linewidth}
        \input{algorithm/superpixel_attack}
    \end{minipage}

\end{figure}

%% file: algorithm/superpixel_attack.tex
\begin{algorithm}[H]
    \caption{\quad Superpixel Attack}
    \label{algorithm:superpixel_attack}
    \begin{algorithmic}[1]
        \REQUIRE{
        Image height $H\in \mathbb{N}$,
        Image width $W\in \mathbb{N}$,
        Number of color channels $C\in \mathbb{N}$,\\
        Allowed perturbation size $\epsilon\in\mathbb{R}^+$,
        Maximum iterations $T\in \mathbb{N}$,\\
        Original image $x_{org}\in \mathcal{D}$,
        Ground truth label $y\in\{1,\dots,Y\}$,
        Segments ratio $r \in \mathbb{N}$,\\
        Classification model $f: \mathcal{D} \to [0,1]^Y$,
        Loss function $L:[0, 1]^Y\times \{1, \dots, Y\} \to \mathbb{R}$,\\
        Projection function $P: \mathbb{R}^{H\times W\times C} \to \mathcal{D}$
        }
        \ENSURE{Image with best loss $x_{best}$\\$ $}

        \STATE $\mathcal{E}_{best} \gets \{\epsilon\}^{H\times W\times C}$ \qquad Initialize perturbation
        \STATE $\mathcal{A} \gets \{(h,w,c) | h\in[1,H], w\in[1,W], c\in[1,C]\}$ \qquad Entire area of image
        \STATE $S \gets \{\mathcal{A}\}$ \qquad Initialize Update Area
        \STATE $\mathcal{L}_{best} \gets -\infty$ \qquad Best loss
        \STATE $n\gets 1$ \qquad Maximum number of superpixel
        \FOR {$t = 1, 2, \dots, T$}
        \STATE $s\in S$, \quad $S \gets S\setminus \{s\}$ \qquad Randomly extract next Update Area
        \STATE $\mathcal{E}_{best} \to \mathcal{E}_{best}[s]+\mathcal{E}_{best}[\mathcal{A} \setminus s]$, \qquad $\mathcal{E}\gets -\mathcal{E}_{best}[s] + \mathcal{E}_{best}[\mathcal{A} \setminus s]$\\
        \qquad \quad Flip perturbation only in Update Area $s\in S$
        \STATE $\hat{x} \gets P(x_{org} + \mathcal{E})$, \quad $\mathcal{L} \gets L(f(\hat{x}), y)$
        \IF {$\mathcal{L} \geq \mathcal{L}_{best}$ \quad Loss increase \qquad}
        \STATE $\mathcal{L}_{best} \gets \mathcal{L}$, \quad $\mathcal{E}_{best} \gets \mathcal{E}$
        \ENDIF
        \IF{$S=\emptyset$ \quad All areas are searched \qquad}
        \STATE $n \gets n \times r$, \ \ $\mathcal{S} \gets SLIC(x_{org}, n)$ \quad Calculate superpixels by SLIC algorithm
        \STATE $S\gets \mathcal{S} \times \{1, \dots, C\}$ \qquad Divide area into superpixels for each RGB channel
        \ENDIF
        \ENDFOR
        \STATE$x_{best}\gets P(x_{org}+\mathcal{E}_{best})$
    \end{algorithmic}
\end{algorithm}

%% file: table/result.tex
\begin{table}[htbp]
    \caption{Comparison experiments with baselines}
    \label{table:result}
    \centering
    \scalebox{0.88}{
        \begin{tabular}{c|c|ccccc|c}
            \multicolumn{2}{c}{100 iter}   & \multicolumn{5}{c}{Attack Success Rate (\%)}                                                                       \\
            \hline

            source                         & Architecture                                 & Parsimon & Square & SignH      & AccSignH & \bf{Superpixel} & diff  \\
            \hline\hline

            Wong\cite{wong}                & ResNet-50                                    & 48.32    & 49.10  & 50.86      & 49.48    & \bf{53.86}      & 3.00  \\
            Engstrom\cite{engstrom}        & ResNet-50                                    & 42.40    & 41.68  & 42.92      & 42.08    & \bf{45.26}      & 2.34  \\
            Salman\cite{salman}            & ResNet-50                                    & 41.24    & 40.42  & 41.98      & 41.06    & \bf{44.44}      & 2.46  \\
            Salman                         & ResNet-18                                    & 52.08    & 51.50  & 52.58      & 52.06    & \bf{56.06}      & 3.48  \\
            Salman                         & WideResNet-50-2                              & 36.84    & 35.64  & 37.82      & 36.54    & \bf{39.84}      & 2.02  \\
            PyTorch$^1$                    & ResNet-50                                    & 33.92    & 47.56  & \bf{50.08} & 38.80    & 47.52           & -2.52 \\
            Debenedetti\cite{debenedetti}  & XCiT-S12                                     & 31.72    & 30.66  & 32.36      & 31.64    & \bf{33.86}      & 1.50  \\
            Debenedetti                    & XCiT-M12                                     & 30.36    & 29.38  & 31.06      & 30.14    & \bf{32.84}      & 1.78  \\
            Debenedetti                    & XCiT-L12                                     & 30.12    & 29.58  & 30.66      & 29.94    & \bf{32.32}      & 1.66  \\
            Singh\cite{singh}              & ViT-S+ConvStem                               & 31.16    & 30.10  & 31.40      & 30.92    & \bf{33.48}      & 2.08  \\
            Singh                          & ViT-B+ConvStem                               & 27.12    & 26.40  & 27.56      & 26.68    & \bf{29.22}      & 1.66  \\
            Singh                          & ConvNeXt-T+ConvStem                          & 30.52    & 29.76  & 30.64      & 30.04    & \bf{32.78}      & 2.14  \\
            Singh                          & ConvNeXt-S+ConvStem                          & 29.26    & 28.46  & 29.72      & 28.98    & \bf{31.34}      & 1.62  \\
            Singh                          & ConvNeXt-B+ConvStem                          & 26.90    & 26.20  & 27.38      & 26.82    & \bf{28.86}      & 1.48  \\
            Singh                          & ConvNeXt-L+ConvStem                          & 25.36    & 24.82  & 25.94      & 25.34    & \bf{26.94}      & 1.00  \\
            Liu\cite{liu}                  & ConvNeXt-B                                   & 26.48    & 25.88  & 26.84      & 26.44    & \bf{28.36}      & 1.52  \\
            Liu                            & ConvNeXt-L                                   & 25.08    & 24.26  & 25.78      & 24.90    & \bf{26.88}      & 1.10  \\
            Liu                            & Swin-B                                       & 26.86    & 26.06  & 27.20      & 26.74    & \bf{28.88}      & 1.68  \\
            Liu                            & Swin-L                                       & 24.16    & 23.36  & 24.62      & 23.80    & \bf{26.06}      & 1.44  \\

            \midrule

            \multicolumn{2}{c}{}                                                                                                                                \\
            \multicolumn{2}{c}{1,000 iter} & \multicolumn{5}{c}{Attack success rate (\%)}                                                                       \\
            \hline

            source                         & Architecture                                 & Parsimon & Square & SignH      & AccSignH & \bf{Superpixel} & diff  \\
            \hline\hline

            Wong                           & ResNet-50                                    & 56.62    & 56.62  & 52.46      & 50.34    & \bf{59.96}      & 3.34  \\
            Engstrom                       & ResNet-50                                    & 48.92    & 48.16  & 45.10      & 44.12    & \bf{51.84}      & 2.92  \\
            Salman                         & ResNet-50                                    & 46.96    & 46.70  & 44.06      & 43.08    & \bf{50.16}      & 3.20  \\
            Salman                         & ResNet-18                                    & 58.60    & 58.72  & 54.92      & 54.26    & \bf{61.98}      & 3.26  \\
            Salman                         & WideResNet-50-2                              & 42.94    & 42.22  & 39.66      & 38.32    & \bf{44.86}      & 1.92  \\
            PyTorch                        & ResNet-50                                    & 72.04    & 84.64  & 80.80      & 55.80    & \bf{87.28}      & 2.64  \\
            Debenedetti                    & XCiT-S12                                     & 37.44    & 36.48  & 33.74      & 32.96    & \bf{39.66}      & 2.22  \\
            Debenedetti                    & XCiT-M12                                     & 36.04    & 35.10  & 32.70      & 31.86    & \bf{37.64}      & 1.60  \\
            Debenedetti                    & XCiT-L12                                     & 35.32    & 34.64  & 32.38      & 31.52    & \bf{37.02}      & 1.70  \\
            Singh                          & ViT-S+ConvStem                               & 35.50    & 35.30  & 32.68      & 32.44    & \bf{37.58}      & 2.08  \\
            Singh                          & ViT-B+ConvStem                               & 31.02    & 30.38  & 28.76      & 28.20    & \bf{32.56}      & 1.54  \\
            Singh                          & ConvNeXt-T+ConvStem                          & 34.88    & 34.50  & 32.24      & 31.58    & \bf{37.12}      & 2.24  \\
            Singh                          & ConvNeXt-S+ConvStem                          & 33.36    & 32.80  & 30.94      & 30.62    & \bf{35.28}      & 1.92  \\
            Singh                          & ConvNeXt-B+ConvStem                          & 30.78    & 30.14  & 28.62      & 28.24    & \bf{32.44}      & 1.66  \\
            Singh                          & ConvNeXt-L+ConvStem                          & 29.24    & 28.80  & 27.30      & 26.42    & \bf{30.64}      & 1.40  \\
            Liu                            & ConvNeXt-B                                   & 30.32    & 29.76  & 28.22      & 27.62    & \bf{31.94}      & 1.62  \\
            Liu                            & ConvNeXt-L                                   & 28.92    & 28.34  & 26.86      & 26.40    & \bf{30.20}      & 1.28  \\
            Liu                            & Swin-B                                       & 30.88    & 30.44  & 28.64      & 28.08    & \bf{32.60}      & 1.72  \\
            Liu                            & Swin-L                                       & 28.18    & 27.44  & 25.86      & 25.32    & \bf{29.90}      & 1.72  \\
            \hline
            \multicolumn{6}{l}{$^1 $ \url{https://pytorch.org/vision/stable/models.html}}
        \end{tabular}
    }
\end{table}

%% file: figure/plot_acc.tex
\begin{figure}[htbp]
    \centering
    \includegraphics[width=0.8\linewidth]{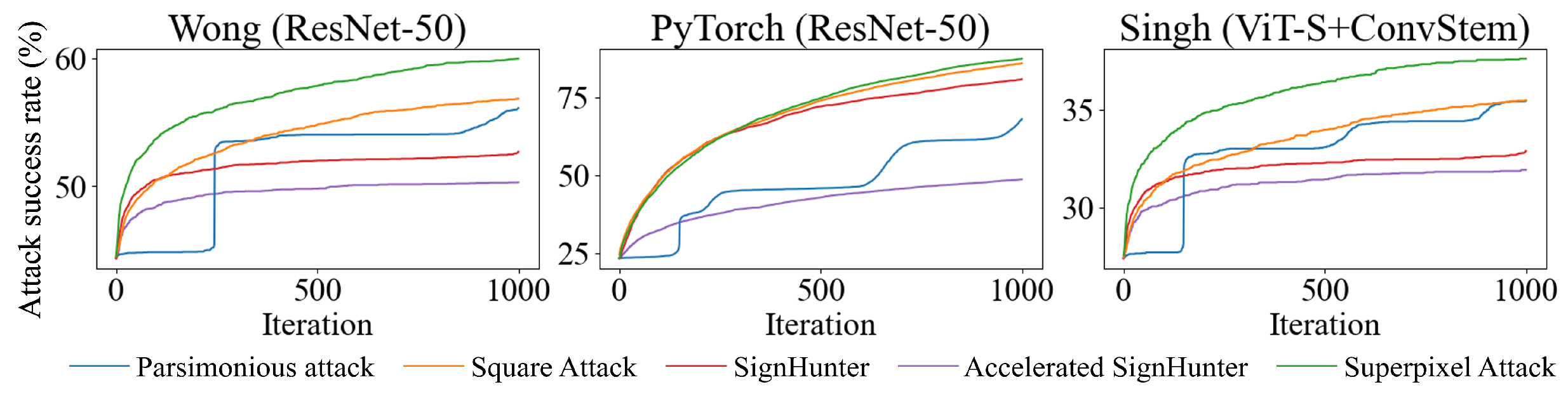}
    \caption{Transition of attack success rates of each attack method}
    \label{figure:plot_acc}
\end{figure}

%% file: figure/computational_time.tex
\begin{figure}[htbp]
    \centering
    \includegraphics[width=0.78\linewidth]{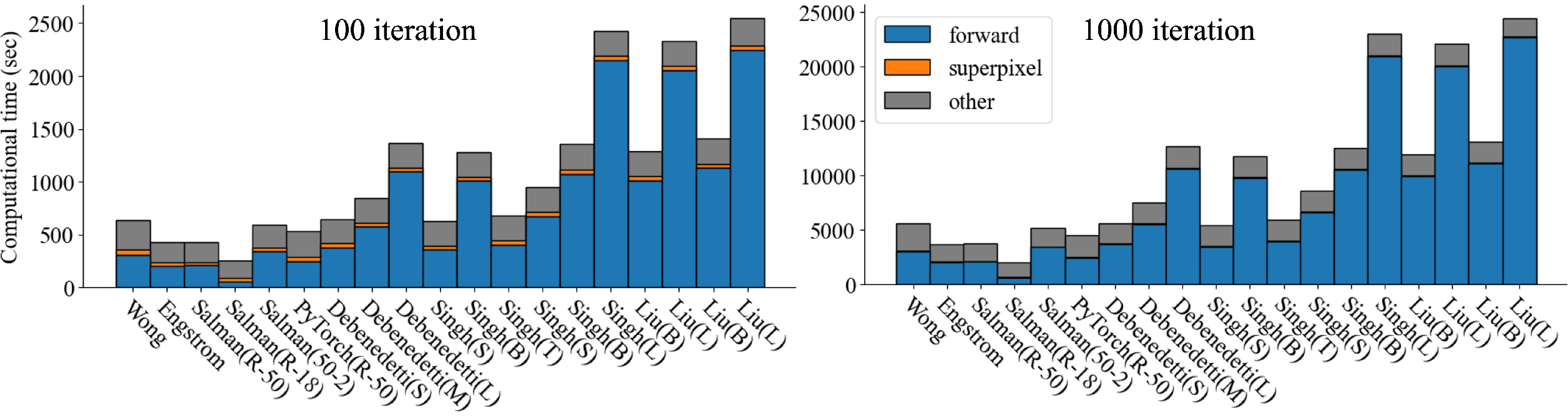}
    \caption{Computational time of superpixels and forward propagation}
    \label{figure:computational_time}
\end{figure}